\documentclass[aps,prapplied,reprint,twocolumn,superscriptaddress]{revtex4-2}

\usepackage{amsmath} % include math package
\usepackage{amssymb} 
\usepackage{graphicx}% Include figure files
\usepackage{dcolumn}% Align table columns on decimal point
\usepackage{bm}% bold math
\usepackage[hidelinks]{hyperref}
\usepackage{color} % for colored text
\usepackage{siunitx}
\usepackage{comment}
\usepackage[capitalise]{cleveref}  % removed [capitalise] option
\crefname{section}{Sec.}{Secs.}%
\usepackage{soul,xcolor}
\usepackage[normalem]{ulem}

\usepackage[nearskip=0pt,farskip=0pt,captionskip=0pt,caption=false]{subfig}
\newcommand{\phantomsubfloat}[1]{{
    \captionsetup[subfigure]{labelformat=empty}
    ~\\[-1.6em]
    \subfloat[][]{#1}
}}

\crefrangelabelformat{figure}{#3#1#4--#5(\crefstripprefix{#1}{#2}#6}
\crefmultiformat{figure}{Figs.~#2#1\xdef\mycreffirstarg{#1}#3}{ and~#2({\crefstripprefix{\mycreffirstarg}{#1}}#3}{, #2({\crefstripprefix{\mycreffirstarg}{#1}}#3}{ and~#2({\crefstripprefix{\mycreffirstarg}{#1}}#3}

\newcommand{\blue}[1]{{\color{blue} #1}}

\newcommand{\Fig}[2][]{\cref{fig:#2}\if #1\empty\else(#1)\fi}

\newcommand\LDst{\bgroup\markoverwith{\blue{\rule[0.5ex]{2pt}{0.4pt}}}\ULon}

\newcommand{\affilTracy}{Institut f\"ur Experimentalphysik, Universit\"at Innsbruck, Technikerstrasse 25, 6020 Innsbruck, Austria}
\newcommand{\affilPhotonics}{Photonics Laboratory, ETH Z\"{u}rich, 8093 Z\"{u}rich, Switzerland}
\newcommand{\affilQC}{Quantum Center, ETH Z\"urich, Z\"urich, Switzerland}

\begin{document}

\title{Hybrid Paul-optical trap with large optical access for levitated optomechanics}

\author{Eric Bonvin}
\author{Louisiane Devaud}
\author{Massimiliano Rossi}
\author{Andrei Militaru}
\affiliation{\affilPhotonics}
\affiliation{\affilQC}

\author{Lorenzo Dania}
\author{Dmitry S. Bykov}
\author{Markus Teller}
\affiliation{\affilTracy}

\author{Tracy E. Northup}
\affiliation{\affilTracy}

\author{Lukas Novotny}
\author{Martin Frimmer}
\affiliation{\affilPhotonics}
\affiliation{\affilQC}

\date{\today}

\begin{abstract}

We present a hybrid trapping platform that allows us to levitate a charged nanoparticle in high vacuum using either optical fields, radio-frequency fields, or a combination thereof. 
Our hybrid approach combines an optical dipole trap with a linear Paul trap while maintaining a large numerical aperture (0.77~NA). 
We detail a controlled transfer procedure that allows us to use the Paul trap as a `safety net' to recover particles lost from the optical trap at high vacuum. 
The presented hybrid platform adds to the toolbox of levitodynamics and represents an important step towards fully controllable `dark' potentials, providing control in the absence of decoherence due to photon recoil.
\end{abstract}

\maketitle

\section{Introduction}
\label{sec:introduction}

Since their invention in the 1970s \cite{ashkin1970acceleration}, optical traps have become indispensable tools in diverse fields ranging from biology to quantum information processing to optomechanics~\cite{bustamante2021optical,Saffman2016,ashkin1997optical}, with the recent emergence of the field of levitodynamics~\cite{millen2019optomechanics,gonzalez-ballestero2021levitodynamics}.
Thanks to their high oscillation frequencies, good isolation from the environment, and large detection efficiency, optical traps have been used to trap and cool the center-of-mass (CoM) motion of nano-sized dielectric particles~\cite{gieseler2012subkelvin}, even reaching the ground state of motion~\cite{delic2020cooling,tebbenjohanns2021quantum,magrini2021realtime,piotrowski2023simultaneous}.
However, since optical traps rely on strongly focused laser beams, they are characterized by small trapping volumes~\cite{gieseler2012subkelvin} and relatively shallow potentials, typically only a few $k_BT_\text{R}$ deep for nanoparticles~(with $k_B$ the Boltzmann constant and $T_\text{R}$ room temperature)\cite{dania2021optical}. 
As a result, even a small unwanted disturbance of the experiment can lead to a loss of the particle from the optical trap, ending the experiment and requiring loading and characterization of a new particle.
Arguably, particle loss is haunting all researchers facing the reality of levitated optomechanics experiments, even though the issue is largely excluded from discussions in the literature as a mere nuisance to those doing the work at the optical table.
Importantly, however, for future studies requiring many experimental realizations of a protocol with the same particle~\cite{gonzalez-ballestero2021levitodynamics}, particle loss will likely be a true show-stopper.

As a levitation technology alternative to optical traps, charged particles can be confined in Paul traps, which are generated by combining radio-frequency (RF) and static (DC) electric fields \cite{paul1990electromagnetic}.
Paul traps were originally applied to suspending atomic and molecular ions, but their use has since been extended to micro- and nano-particle systems~\cite{Xiong2019}.
For nanoparticles, in contrast to optical traps, Paul traps generate large (typically millimeters in size) and deep trapping potentials  (exceeding 1000~$k_BT_\text{R}$), but typically have lower trapping frequencies~\cite{bykov2019direct,dania2021optical}.
This observation initiated the development of hybrid Paul-optical traps in the levitodynamics community, combining the advantages of both methods.
The concept of hybrid trapping has been developed in the atomic and ion physics community to investigate ensembles of neutral atoms and ions and to optically trap ions~\cite{Schaetz2017}. 
For levitodynamics, it would be attractive to surround a stiff but small optical trap with the large but loose potential of a Paul trap. 
In this fashion, experiments can be performed in the optical trap, with the Paul trap acting as a safety net should the particle escape the optical potential. The same particle can then be recovered in the optical trap for further experiments.

Inspiration for hybrid Paul-optical traps can be drawn from two lines of prior research. First, designs facilitating spectroscopy on absorbing objects in levitation have been developed~\cite{nagornykh2015cooling,conangla2018motion,conangla2020extending,delord2017diamonds}. These approaches are essentially Paul traps equipped with limited optical access.
Second, truly hybrid approaches, allowing for both optical and Paul trapping, have been demonstrated~\cite{bykov2022hybrid, millen2015cavity,fonseca2016nonlinear}. Here, limited optical access was alleviated by using counter-propagating optical fields in free space or optical resonators.
For both approaches, the geometric difficulty of combining the two confinement techniques leads to compromises in one or the other trapping method---either the Paul trap electrodes are located far from the particle, producing low Paul trap frequencies, or they restrict optical access, which reduces the particle detection efficiency~\cite{tebbenjohanns2019optimal} and makes optical trapping more challenging.
These trade-offs limit the usefulness of Paul traps as safety nets at the present stage.
Specifically, particle recovery from the Paul to the optical trap, even when active feedback is applied, currently relies on damping from a buffer gas, restricting hybrid operation to low vacuum, with a lowest reported pressure of \SI{5e-2}{\milli \bar}~\cite{bykov2022hybrid}.

In this paper, we describe a hybrid trap that consists of a single beam optical trap with high-numerical-aperture (NA) suitable for measurement-based quantum control, overlapped with a Paul trap that does not compromise optical access.
The Paul trap captures the particle, should it escape from the optical trap.
Feedback cooling of the particle motion in the Paul trap allows us to recover the particle back into the optical trap at pressures down to \SI{3e-{6}}{\milli\bar}.

%%% Fig. Setup %%%%%%%%%%%%%%%%%%%%%%%%%%%%%%%%%%%%%%%%
\begin{figure}[t]
    \includegraphics[width=1\columnwidth]{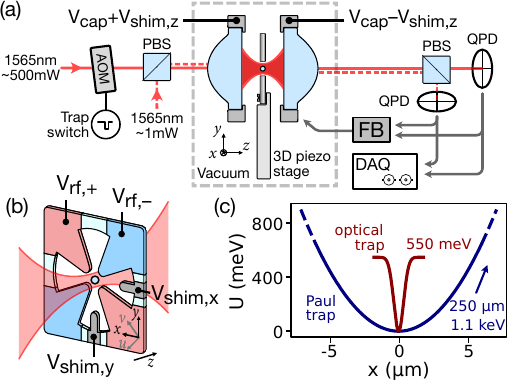}
    \phantomsubfloat{\label{fig:setup}}
    \phantomsubfloat{\label{fig:electrodes}}
    \phantomsubfloat{\label{fig:potentials}}
    \caption{
    (a)~Schematic diagram of the hybrid trapping setup: a trapping laser passes through an acousto-optic modulator (AOM) before being focused through a high-NA lens to form an optical trap in vacuum. A weaker (cross-polarized) measurement beam is overlapped with the trapping beam using a polarizing beam splitter (PBS). The forward-scattered light of both beams is split by a second PBS and sent onto separate quadrant photodiodes (QPD) for detection.
    The signals from both QPDs can selectively be used for feedback cooling (FB) and/or recorded via a data-acquisition module (DAQ).
    A microfabricated wheel trap is held in the focal plane of the trapping lens by a 3-axis piezo stage. 
    (b)~Diagram of the wheel trap electrode geometry. Two pairs of RF electrodes (blue and red) are driven by high-voltage RF tones ($V_{\text{rf},+}$, $V_{\text{rf},-}$)  with opposite phases. Two shim electrodes are used to apply bias voltages $V_{\text{shim},x}$ and $V_{\text{shim},y}$ to generate Coulomb forces along the $x$ and $y$ directions, respectively.
    (c)~Plot of typical trapping potentials: 
    The stiff optical trap is much shallower than the loose Paul trap.
    }
    \label{fig:setup_full_figure}
\end{figure}
%%%%%%%%%%%%%%%%%%%%%%%%%%%%%%%%%%%%%%%%%%%%%%%%%%%

\section{Experimental setup}
A schematic of the experimental setup is shown in~\cref{fig:setup}.
 Let us discuss the separate building blocks.

\paragraph*{Optical trap.}
A laser beam (\SI{500}{\milli \watt}, \SI{1565}{\nano \meter}) polarized along $x$ (hereafter trapping beam) is focused by a lens (0.77~NA, effective focal length \SI{3.3}{\milli \meter}, working distance \SI{1.6}{\milli \meter}) to form the optical trap, in which we levitate silica nanospheres (diameter \SI{177}{\nano\meter}).
Typical oscillation frequencies in the optical trap along the $(x,y,z)$ directions are $2\pi\times(69,74,15)$~kHz, respectively. 
The optical trap can be switched on and off with an acousto-optic modulator (AOM).

\paragraph*{Loading.}
We load the optical trap at ambient pressure by creating an aerosol of a solution of particles suspended in isopropanol close to the focus of the trapping lens \cite{summers2008trapping}.
As we decrease the gas pressure, an optically trapped silica nanoparticle undergoes a sudden change in its properties, which manifests itself as a drop in oscillation frequencies, as well as an increase in charge~\cite{ricci2022chemical}.
Our particles typically acquire a positive charge-to-mass ratio of \SIrange{4}{5}{\coulomb \per \kilogram}, or approximately 150 elementary charges.
We rely on this charge to make use of the Paul trap.

\paragraph*{Paul trap.}
The Paul trap, illustrated in Fig.~\ref{fig:electrodes}, consists of a \SI{300}{\micro \meter} thick fused silica substrate micro-machined and gold-coated (FEMTOprint, Switzerland) to have six electrodes: two pairs of RF electrodes generate a quadrupole potential for in-plane confinement, and two shim electrodes are designed to apply small in-plane electric fields.
This geometry is commonly referred to as a wheel trap \cite{chen2017ticking,chen2017sympathetic,brewer2019al+,teller2023integrating}.
In our case, the tips of the RF electrodes are $\SI{250}{\micro \meter}$ away from the stable trapping point.
Our wheel trap's high optical access (up to $0.85~\text{NA}$), allows for passing a strongly focused optical beam without obstruction.
The orientation of the RF electrodes defines two new axes $u$ and $v$ that are tilted by 45$^\circ$ with respect to the $x$ and $y$ axes.
Out-of-plane confinement (and stray field compensation along $z$) is achieved by applying DC voltages $V_\text{cap}$ to the lens holders, which double as endcap electrodes, located a distance $\SI{1.6}{\milli\meter}$ away from the trap center.
For position control relative to the optical trap, the wheel trap is held in the focal plane of the trapping lens by a 3-axis nanopositioner. The RF electrodes are driven by RF signals with opposite phase, angular frequency $\Omega_\text{d}$ and amplitude $V_\text{rf}$.
These values control the oscillation frequency of the $u$ and $v$ modes of the Paul trap.
A small DC offset $V_\text{off}$ is added to one pair of RF electrodes and subtracted from the other to spectrally separate the $u$ and $v$ modes.
Three additional adjustable voltages, $V_{\text{shim,}x}$, $V_{\text{shim,}y}$, and $V_{\text{shim,}z}$, are used to generate DC forces along $x$, $y$, and $z$, respectively.
While $V_{\text{shim,}x}$ and $V_{\text{shim,}y}$ are applied to their respective electrodes on the wheel trap, $V_{\text{shim,}z}$ is added to one endcap electrode and subtracted from the other.
Typical values for our experiment are $\Omega_\text{d}=2\pi\times\SI{33}{\kilo \hertz}$, $V_\text{rf}=\SI{200}{\volt}$, $V_\text{off}=\SI{5}{\milli\volt}$, and $V_\text{cap}=\SI{70}{\volt}$, which leads to oscillation frequencies in the Paul trap along the $u,v,z$ directions of $2\pi\times (5.5, 6, 3)$~kHz, respectively.
With our system and particles, we can access trap frequencies in a range of \SIrange{0.1}{10}{\kilo \hertz}.
This range is restricted at the lower end by insufficient trap depth, and at the upper end by the maximum available voltages and the particle's charge-to-mass ratio.
Due to the high voltages required to drive the Paul trap, we only operate it below \SI{e-2}{\milli \bar} to avoid arc discharges~\cite{paschen1889ueber}.

\paragraph*{Comparison of potentials.}
Figure~\ref{fig:potentials} highlights the differences between the two trapping potentials.
The optical trap is much stiffer, but also much shallower than the Paul trap.
At a depth of roughly \SI{550}{\milli \electronvolt}, particle escape is quite frequent if the particle motion is driven to large oscillation amplitudes, for example due to an incorrectly set phase in a feedback control loop.
In contrast, at a depth around \SI{1.1}{\kilo \electronvolt}, the looser Paul trap is much deeper than the optical trap.
Thus, as long as the charge-to-mass ratio remains in the stable regime, a particle trapped in the Paul trap will virtually never escape.

\paragraph*{Detection.}
When the particle is held in the optical trap, the forward-scattered light of the trapping beam is collected by a lens identical to the trapping lens and sent onto a quadrant photodiode (QPD) for interferometric detection~\cite{gittes1998interference}.
To monitor the position of the particle in the Paul trap in the absence of the trapping beam, a second, much weaker laser beam (\SI{1}{\milli \watt}, \SI{1565}{\nano \metre}, hereafter called measurement beam) polarized along $y$ is overlapped with the trapping beam via a polarizing beam splitter (PBS).
In the forward direction, light from the measurement beam is directed onto a separate QPD with a second PBS. 

\paragraph*{Feedback cooling.}
In order to stabilize the particle motion, particularly at low pressures, we apply feedback cooling to the particle along all three spatial dimensions.
The particle's position signals are processed with digital bandpass filters implemented on field-programmable gate arrays, to generate cold damping signals~\cite{Poggio2007}. We adapt the spectral locations of the filters depending on whether the particle is trapped optically or electrically. These feedback signals are applied as voltages to the corresponding shim electrodes for 3D feedback cooling~\cite{tebbenjohanns2019cold, conangla2019optimal}.

%%% Fig. alignment  %%%%%%%%%%%%%%%%%%%%%%%%%%%%%%%%%%%%%%%%
\begin{figure}[t]
    \includegraphics[width=1\columnwidth]{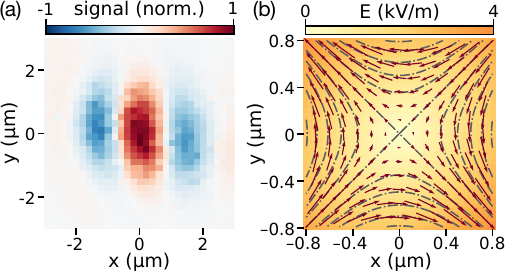}
    \phantomsubfloat{\label{fig:detection_pattern}}
    \phantomsubfloat{\label{fig:electric_field_potential}}
    \caption{
    \label{fig:alignment}
    (a)~2D scan of the local $x$ position sensitivity of the measurement beam.
    The maximum in this map corresponds to the focus of the beam.
    (b)~Measured electric field generated by the RF electrodes of the wheel trap (red arrows, orange background). The measured field closely matches that expected from a normalized quadrupole potential (gray dash-dotted lines).
    }
\end{figure}
%%%%%%%%%%%%%%%%%%%%%%%%%%%%%%%%%%%%%%%%%%%%%%%%%%%

\paragraph*{Characterization of optical detection.}
To characterize the optical detection scheme, we levitate a particle in the Paul trap with the measurement beam enabled and disable the optical trap.
We apply a small modulation tone to the $x$ shim electrode, thereby driving the particle along $x$. We detect the response of the particle to this drive tone by demodulating the $x$ channel of the QPD signal at the drive-tone frequency. In this fashion, we measure the local position sensitivity along $x$ of our measurement beam.
By performing a 2D scan of the piezo stage in the $xy$ plane (which means that we effectively scan the particle by moving the Paul trap chip), we are able to map the local $x$ position sensitivity, shown in~\cref{fig:detection_pattern}.
The signal is maximal at the center of the detection pattern, which corresponds to the focal point of the measurement beam.
Since the trapping and measurement beams are overlapped, we can use this procedure to align both trapping potentials.
Furthermore, we observe that the sign of the local $x$ position sensitivity is inverted when the particle is displaced along $x$ by a distance comparable to the wavelength, which is a result of the interference of the field scattered by the particle with the measurement beam acting as a reference field on the detector~\cite{gittes1998interference,dania2021optical}.

\paragraph*{Characterization of electric potential.}
Conversely, we can make use of the optical trap to characterize the electric potential. With the particle held in the optical trap, we enable the RF drive of the Paul trap and demodulate the detector signals generated by the measurement beam on the $x$ and $y$ channel of the QPD detector. After calibrating the detector signal using the procedure outlined in~\cite{hebestreit2018calibration}, we can use the particle as a field sensor~\cite{Frimmer2017}.
By performing a 2D scan of the piezo stage in the $xy$-plane, moving the RF field distribution relative to the particle held in the tight optical focus, we record a snapshot of the RF field, shown as red arrows in~\cref{fig:electric_field_potential}. Our measurement matches well with a quadrupolar field distribution, shown as black dashed lines in~\cref{fig:electric_field_potential}, as expected for our electrode geometry.
We note that a point of particular interest in this map is the RF null, i.e., the point where the RF field is minimal (and even vanishes, for an ideal electrode geometry). At this location, a particle in the optical trap is minimally affected by the Paul trap's RF field.

\paragraph*{Stray field compensation.}
When the particle is held in the Paul trap, its equilibrium position may differ from the RF null due to stray electric fields. In order to compensate for these forces, we adjust the shim voltages of the trap's electrodes until the measured micromotion amplitude is minimized. In fact, this process compensates for all DC forces displacing the particle, including gravity.

\section{Paul trap as a safety net}

\paragraph*{Operation.}
To operate the Paul trap as a safety net, we keep it enabled while the particle is trapped optically, and carefully align the optical trap to the RF null of the Paul trap by minimizing the micromotion amplitude.
% We carefully align the optical trap to the RF null of the Paul trap, by minimizing the micromotion amplitude, as outlined in the discussion of~\cref{fig:electric_field_potential}.
When the particle escapes from the optical trap (which we provoke in our experiments by turning off the trapping beam), it is caught by the Paul trap, at which point we start our recovery procedure to transfer the particle back to the optical trap.

\begin{figure}[t]
    \label{fig:transfer_success_fail}
    \includegraphics[width=1\columnwidth]{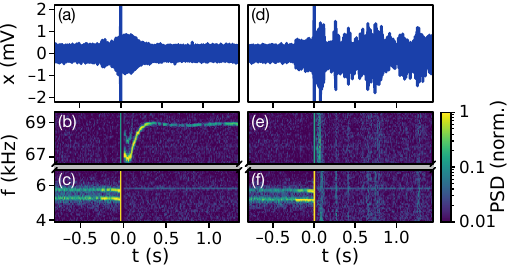}
    \phantomsubfloat{\label{fig:transfer_success_time_trace}}
    \phantomsubfloat{\label{fig:transfer_success_optical_spectrum}}
    \phantomsubfloat{\label{fig:transfer_success_paul_spectrum}}
    \phantomsubfloat{\label{fig:transfer_fail_time_trace}}
    \phantomsubfloat{\label{fig:transfer_fail_optical_spectrum}}
    \phantomsubfloat{\label{fig:transfer_fail_paul_spectrum}}
    \caption{
    (a)~Time trace on $x$ detector of measurement beam during a successful particle transfer from the Paul to the optical trap. The particle is initially feedback cooled in the Paul trap ($t<0$). At $t=0$, the optical trap is switched on.
    The PSD of this time trace is shown in the spectral range of the $x$ mode in the optical trap in (b) and the spectral range of the $u$ and $v$ modes in the Paul trap in (c). 
    Each PSD is calculated from a \SI{17.5}{\milli\second} segment of the time trace sampled at \SI{234}{\kilo\hertz}.
    (d)~Time trace of a failed transfer attempt together with its PSD in (e) and (f).
    Measurements are taken at \SI{6e-5}{\milli\bar}.
    }
\end{figure}

\paragraph*{Particle recovery.}
We initialize the particle recovery procedure by disabling the optical trap, enabling the measurement beam, and adapting our feedback cooling parameters to match the oscillation frequencies of the particle in the Paul trap.
Additionally, we ensure that the RF potential is aligned with the optical potential by maximizing the particle response to an external drive tone, centering the particle in the optical detection pattern as described in the discussion of~\cref{fig:detection_pattern}.
We then execute the following sequence: (i)~feedback cooling at Paul trap frequencies is disabled, (ii)~the optical trap is switched on, (iii)~feedback cooling is enabled at optical trap frequencies, and (iv) the success of the transfer is evaluated. 
For this final step, we analyze the particle position recorded by the measurement beam during the transfer attempt. An example from  a successful transfer is shown in \cref{fig:transfer_success_time_trace}.
The optical trap is switched on at time $t=0$, which causes an abrupt spike in the measured signal, followed by an increase in signal amplitude that decays after about 0.25~s. The success of the transfer becomes visible when analyzing the timetrace from~\cref{fig:transfer_success_time_trace} in the spectral domain. To this end, we investigate the power spectral density (PSD) of the signal in the vicinity of the optical trap frequencies [\cref{fig:transfer_success_optical_spectrum}] and Paul trap frequencies [\cref{fig:transfer_success_paul_spectrum}].
Before the transfer ($t<0$), we recognize peaks at the $u$ and $v$ modes in the Paul trap between 5 and 6~kHz, indicating the presence of the particle in the Paul trap.
After the transfer ($t>0$), these modes vanish and the $x$ mode of the particle in the optical trap appears around 69~kHz. This observation indicates successful transfer of the particle into the optical trap.
We observe a frequency increase from 67 to 69~kHz during the first 0.25~s after particle recapture, coinciding with the activation of feedback cooling. We attribute this frequency change to the Duffing non-linearity of the optical potential~\cite{Gieseler2014nonlinear}.

For comparison, \cref{fig:transfer_fail_time_trace} shows an example of a failed transfer, together with the corresponding spectrograms in~\cref{fig:transfer_fail_optical_spectrum,fig:transfer_fail_paul_spectrum}. 
After a failed transfer, the particle is still present in the Paul trap, but no clear spectral signature from either trap is present in the detector signal. The reason is that the particle's oscillation amplitude is much larger than the detection beam.
From an energetic perspective, one might expect that an excited particle would eventually thermalize with the surrounding gas and fall into the optical trap.
This is indeed the case at pressures above~\SI{e-2}{\milli\bar} [see discussion of~\cref{fig:transfer-success-pressure} below]. However, below~\SI{e-3}{\milli\bar}, we observe that the particle never settles into the optical potential, even on timescales exceeding the damping time.
We speculate that, at low pressures, non-conservative optical forces (such as the scattering force) add energy to the particle at a rate exceeding the damping rate provided by the gas.
Indeed, radiation pressure in an optical tweezer is a non-conservative force, which can cause the particle to follow a trajectory that adds mechanical energy to the system~\cite{ashkin1992forces, roichman2008optical, roichman2008influence}.

%%% Fig. transfer traces  %%%%%%%%%%%%%%%%%%%%%%%%%%%%%%%%%%%%%%%%
%%%%%%%%%%%%%%%%%%%%%%%%%%%%%%%%%%%%%%%%%%%%%%%%%%%

\section{Key parameters for successful recovery}

Based on the above observations, we can increase the likelihood of successful transfer by controlling two parameters: the mean oscillation amplitude of the particle in the Paul trap (via feedback cooling), and the relative position of the Paul and optical traps (via trap alignment). We proceed by investigating the dependence of the transfer success rate on these two parameters.

\begin{figure}[t]
    \includegraphics[width=1\columnwidth]{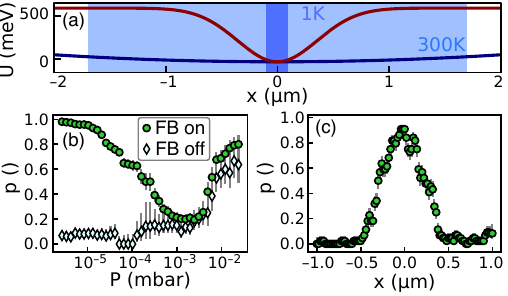}
    \phantomsubfloat{\label{fig:potentials-zoomed}}
    \phantomsubfloat{\label{fig:transfer-success-pressure}}
    \phantomsubfloat{\label{fig:transfer-success-alignment}}
    \caption{
    (a)~Comparison of the calculated optical (red) and Paul trap (blue) potentials $U$ as a function of position $x$. The area shaded in light blue indicates the 95\% confidence interval of the particle's location at room temperature (\SI{300}{\kelvin}) in the Paul trap. Dark blue shows the corresponding confidence interval under feedback cooling to \SI{1}{\kelvin} in the Paul trap. 
    (b)~Measured transfer success rate versus pressure with (green circles) and without (blue diamonds) feedback cooling. Below \SI{e-3}{\milli\bar}, feedback cooling becomes effective and increases the likelihood of successful transfer.
    (c)~Measured transfer success rate versus  displacement of the Paul trap relative to the optical trap. As the particle is moved away from the waist of the optical beam, successful transfers become less likely.
    }
    \label{fig:transfer_success_with_pressure}
\end{figure}

\paragraph*{Feedback cooling.}
For our system parameters, the thermal oscillation amplitude of a particle in the Paul trap with a CoM temperature of \SI{300}{\kelvin} is \SI{850}{\nano\meter}, a dimension comparable to the size of the optical trap.
The particle therefore spends a significant amount of time outside of the optical trap, making successful transfers less likely.
In contrast, when feedback cooled to a CoM temperature of \SI{1}{\kelvin} in the Paul trap, the RMS oscillation amplitude becomes \SI{50}{\nano\meter}, effectively confining the particle motion to within the beam waist.
To illustrate this point, \cref{fig:potentials-zoomed} shows the optical and Paul trap potentials, as well as the 95\% confidence intervals of the particle's instantaneous position in the Paul trap at 1~K and at 300~K.

As an experimental verification of this dependence, at a fixed pressure, we repeatedly attempt transfers from the Paul to the optical trap, and record the success rate. In~\cref{fig:transfer-success-pressure}, we plot the success rate as a function of pressure, both in the presence and in the absence of feedback cooling in the Paul trap, for a total of 743 transfer attempts.
For this experiment, feedback gain is adjusted once at \SI{3e-6}{\milli\bar} and held constant throughout all measurements. Since CoM temperature under feedback cooling depends on gas damping, adjusting the gas pressure is an indirect way of controlling the particle's mean oscillation amplitude in the Paul trap~\cite{tebbenjohanns2019cold, conangla2019optimal, dania2021optical}.
In~\cref{fig:transfer-success-pressure}, we observe a transfer success rate above 50\%, both with and without feedback, at pressures above~\SI{e-2}{\milli \bar}, which we attribute to gas damping. At these pressures, feedback cooling is ineffective.
Reducing pressure, we observe a decrease in success rate to below 30\% between \SIlist{4e-3}{\milli \bar} and \SIlist{5e-4}{\milli \bar}, both with and without feedback cooling.  In this pressure regime, the reduced gas damping is insufficient to slow down the particle once the optical trap has been enabled, yet the pressure is still too high for effective feedback cooling, which therefore fails to reduce the mean oscillation amplitude of the particle before the transfer attempt.
At pressures below \SI{e-3}{\milli \bar}, we observe an increase in the success rate of transfers under feedback cooling. Here, gas damping is still absent, but feedback cooling is able to effectively reduce the oscillation amplitude of the particle, localizing it within the waist of the optical beam prior to the transfer. At pressures below \SI{e-5}{\milli\bar}, the transfer success rate is boosted beyond 90\% by feedback control.
We conclude that efficient 3D-feedback cooling in the Paul trap is a key element for a safety net in the underdamped regime.

\paragraph*{Trap alignment.}
Next, we investigate the importance of relative alignment between the Paul and optical traps for successful particle recovery. 
To this end, at a pressure of \SI{e-5}{\milli\bar}, we perform 40 transfer attempts with a fixed relative position $x$ of the Paul trap relative to the optical trap and determine the transfer success rate. We then sweep the position of the Paul trap with the piezo stage to obtain the success rate as a function of trap position, shown in~\cref{fig:transfer-success-alignment}.
We observe that the success rate is maximized to 90\% when the traps are aligned ($x=0$), but drops to 50\% when the Paul trap is displaced by $x=250~$nm, and below 10\% for $x>500~$nm.  
Thus, we conclude that precise alignment between the optical and the Paul trap is crucial to harness the hybrid trap as a safety net.

\paragraph*{Recovery after failed transfer.}
As mentioned in the discussion of~\cref{fig:transfer_fail_time_trace}, after a failed transfer attempt, the particle motion is driven by the presence of the trapping beam to an amplitude largely exceeding the waist of the measurement beam. As a result, the detected signal is unsuitable for feedback cooling. To regain control over the particle motion after a failed transfer attempt, we switch off the trapping beam and rely on residual gas damping to reduce the oscillation amplitude. 
For context, at \SI{e-5}{\milli\bar}, the signal is usually restored within about \SI{15}{\second}.
Once a signal suitable for feedback cooling in the Paul trap is recovered, the next transfer according to our protocol can be attempted.

The overhead in experiment time generated by the waiting period after a failed transfer attempt leads us to two important conclusions. First, maximizing the transfer success rate by optimizing the feedback cooling and trap alignment is crucial to avoid failed transfer attempts. Second, the waiting (i.e., damping) time, which scales with the inverse of the gas pressure, may become too long to be acceptable in ultra-high vacuum conditions~\cite{dania2023ultrahigh}. 
A logical next development step is therefore a detection mechanism that can provide a position signal suitable for feedback control of a particle on a trajectory several microns from the Paul trap center~\cite{dania2023ultrahigh}. The timescale for recovery of a particle after a failed transfer attempt is then set by the feedback system instead of the gas damping.

%%% Fig. transfer rates  %%%%%%%%%%%%%%%%%%%%%%%%%%%%%%%%%%%%%%%%
%%%%%%%%%%%%%%%%%%%%%%%%%%%%%%%%%%%%%%%%%%%%%%%%%%%

\section{Conclusion}
We have described a hybrid trap which adds a Paul trap to an optical trap without compromising on the latter's performance.
As a specific application, we have demonstrated harnessing the Paul trap as a safety net for a particle that has escaped the optical trap. 
Furthermore, we have shown how careful alignment of the two trapping potentials together with feedback cooling enables the reliable transfer of a particle captured in the Paul trap back to the optical trap at pressures down to \SI{3e-6}{\milli\bar}.

We expect the demonstrated safety-net functionality to become an important technology for future levitodynamics experiments that require repeated experimental runs using the same particle. 
Indeed, having a safety net emboldens one to explore more ambitious experiments that require thousands of repetitions of an experimental protocol without losing the particle.
With more ambitious goals in mind, we envision bringing this technology to higher vacuum and cryogenic environments~\cite{magrini2021realtime,tebbenjohanns2021quantum}.

Beyond only using the Paul trap as a safety net, it is tantalizing to think of integrating it into optomechanical protocols as a control element. For example, the photon-recoil-free nature of the Paul trap potential could be used to measure sources of decoherence beyond photon shot noise, such as black-body radiation due to the trapped particle's internal temperature~\cite{Romero-Isart2011-quantumSuperpos}, or anomalous heating from nearby surfaces~\cite{Brownnutt2015}.
Additionally, the large frequency difference between the optical and Paul traps would allow for protocols to squeeze the levitated nanoparticle's motional state~\cite{janszky1986squeezing,graham1987squeezing}.
Finally, the Paul trap could be adapted to generate more complex potential landscapes, such as those required for rapid state expansion and to generate non-Gaussian states of mechanical motion~\cite{weiss2021large,roda-llordes2023macroscopic}.

\begin{acknowledgments}
This research was supported by the Swiss National
Science Foundation (SNF) through Grants No.\ 212599 and No.\ 217122, NCCR-QSIT program (Grant No.\ 51NF40-160591), the European
Union’s Horizon 2020 research and innovation program
under Grant No.\ 951234 (Q-Xtreme), and by the Austrian Science Fund (FWF) Project No. Y951 and Grant No. I5540. 
\end{acknowledgments}

\bibliography{safety_net_paper}

\end{document}